\documentclass[aps,twocolumn,prapplied,reprint,superscriptaddress,floatfix,longbibliography]{revtex4-2}
\usepackage[utf8]{inputenc}
\usepackage[T1]{fontenc}
\usepackage[colorlinks=true,linkcolor={blue!90!black},citecolor = {blue!90!black},urlcolor={blue!90!black}]{hyperref}
\usepackage{appendix}
\usepackage{amsmath}
\usepackage{amssymb}
\usepackage{graphicx}
\usepackage{afterpage}
\usepackage{amsfonts}
\usepackage{amssymb}
\usepackage{graphicx}
\usepackage{braket}
\usepackage{textcomp}
\usepackage{bm}
\usepackage{soul}
\usepackage[dvipsnames]{xcolor}
 \usepackage{float}
\usepackage{gensymb}
\usepackage{siunitx} 
\usepackage{tikz}
\usepackage{placeins}

\usepackage[]{newtxtext}
\usepackage[subscriptcorrection,nosymbolsc,smallerops,bigdelims]{newtxmath} 
\DeclareMathAlphabet{\mathcal}{OMS}{cmsy}{m}{n}
\DeclareMathAlphabet{\mathbcal}{OMS}{cmsy}{b}{n}

\usepackage{orcidlink}
\newcommand{\orcid}[1]{\orcidlink{#1}}

\usepackage[activate={true,nocompatibility},final,tracking=alltext,kerning=true,spacing=true,protrusion=true,factor=1050,stretch=5,shrink=5,selected=true,letterspace=-0]{microtype}

\begin{document}
\title{Determining Strain Components in a Diamond Waveguide from Zero-Field ODMR Spectra of \texorpdfstring{NV$^{-}$}{} Center Ensembles}

\author{M.~Sahnawaz~Alam\orcid{0000-0001-6599-4964}}
\email[Corresponding author: ]{sahnawaz.alam@pwr.edu.pl}
\affiliation{Institute of Theoretical Physics, Wroc{\l}aw University of Science and Technology, 50-370 Wroc{\l}aw, Poland}

\author{Federico~Gorrini\orcid{0000-0002-6420-4104}}
\email[Corresponding author: ]{federico.gorrini@polito.it}
\affiliation{Department of Molecular Biotechnology and Health Sciences, University of Torino, 10126 Torino, Italy}
\affiliation{Dipartimento Scienza Applicata e Tecnologia, Politecnico di Torino, 10129, Torino, Italy}

\author{Micha{\l}~Gawe{\l}czyk\orcid{0000-0003-2299-140X}}
\email[Corresponding author: ]{michal.gawelczyk@pwr.edu.pl}

\affiliation{Institute of Theoretical Physics, Wroc{\l}aw University of Science and Technology, 50-370 Wroc{\l}aw, Poland}

\author{Daniel~Wigger\orcid{0000-0002-4190-8803}}
\affiliation{Department of Physics, University of M\"unster, 48149, M\"unster, Germany}

\author{Giulio~Coccia\orcid{0000-0002-5095-5038}}
\affiliation{Institute for Photonics and Nanotechnologies (IFN) CNR, 20133 Milano, Italy}

\author{Yanzhao~Guo}
\affiliation{School of Engineering, Cardiff University, Cardiff CF24 3AA, United Kingdom}
\affiliation{Translational Research Hub,Cardiff CF24 4HQ, United Kingdom}

\author{Sajedeh~Shahbazi}
\affiliation{Institute for Quantum Optics, Ulm University, D-89081 Ulm, Germany}
\affiliation{Center for Integrated Quantum Science and Technology (IQst), Ulm University, D-89081 Ulm, Germany}

\author{Vibhav~Bharadwaj\orcid{0000-0002-7974-2597}}
\affiliation{Institute for Photonics and Nanotechnologies (IFN) CNR, 20133 Milano, Italy}
\affiliation{Institute for Quantum Optics, Ulm University, D-89081 Ulm, Germany}
\affiliation{Department of Physics, Indian Institute of Technology Guwahati, 781039 Assam, India}

\author{Alexander~Kubanek\orcid{0000-0002-2417-1985}}
\affiliation{Institute for Quantum Optics, Ulm University, D-89081 Ulm, Germany}
\affiliation{Center for Integrated Quantum Science and Technology (IQst), Ulm University, D-89081 Ulm, Germany}

\author{Roberta~Ramponi\orcid{0000-0002-1806-9706}}
\affiliation{Institute for Photonics and Nanotechnologies (IFN) CNR, 20133 Milano, Italy}

\author{Paul~E~Barclay\orcid{0000-0002-9659-5883}}
\affiliation{Institute for Quantum Science and Technology, University of Calgary, T2N 1N4 Calgary, Canada}

\author{Anthony~J.~Bennett\orcid{0000-0002-5386-3710}}
\affiliation{School of Engineering, Cardiff University, Cardiff CF24 3AA, United Kingdom}
\affiliation{Translational Research Hub,Cardiff CF24 4HQ, United Kingdom}

\author{John~P.~Hadden\orcid{0000-0001-5407-6754}}
\affiliation{School of Engineering, Cardiff University, Cardiff CF24 3AA, United Kingdom}
\affiliation{Translational Research Hub,Cardiff CF24 4HQ, United Kingdom}

\author{Angelo~Bifone\orcid{0000-0002-7423-3419}}
\affiliation{Department of Molecular Biotechnology and Health Sciences, University of Torino, 10126 Torino, Italy}
\affiliation{Center for Sustainable Future Technologies, Istituto Italiano di Tecnologia, 10144 Torino, Italy}

\author{Shane~M.~Eaton\orcid{0000-0003-0805-011X}}
\affiliation{Institute for Photonics and Nanotechnologies (IFN) CNR, 20133 Milano, Italy}

\author{Pawe{\l}~Machnikowski\orcid{0000-0003-0349-1725}}
\affiliation{Institute of Theoretical Physics, Wroc{\l}aw University of Science and Technology, 50-370 Wroc{\l}aw, Poland}

\begin{abstract}
The negatively charged nitrogen-vacancy (NV\(^-\)) center in diamond has shown great potential in nanoscale sensing and quantum information processing due to its rich spin physics. An efficient coupling with light, providing strong luminescence, is crucial for realizing these applications. Laser-written waveguides in diamond promote NV\(^-\) creation and improve their coupling to light but, at the same time, induce strain in the crystal. The induced strain contributes to light guiding but also affects the energy levels of NV\(^-\) centers. We probe NV\(^-\) spin states experimentally with the commonly used continuous-wave zero-field optically detected magnetic resonance (ODMR). In our waveguides, the ODMR spectra are shifted, split, and consistently asymmetric, which we attribute to the impact of local strain. To understand these features, we model ensemble ODMR signals in the presence of strain. By fitting the model results to the experimentally collected ODMR data, we determine the strain tensor components at different positions, thus determining the strain profile across the waveguide. This shows that zero-field ODMR spectroscopy can be used as a strain imaging tool. The resulting strain within the waveguide is dominated by a compressive axial component transverse to the waveguide structure, with a smaller contribution from vertical and shear strain components. 
\end{abstract}

\maketitle
\date{\today}

\section{Introduction}

\begin{figure*}[ht]
\centering
\includegraphics[width=\linewidth]{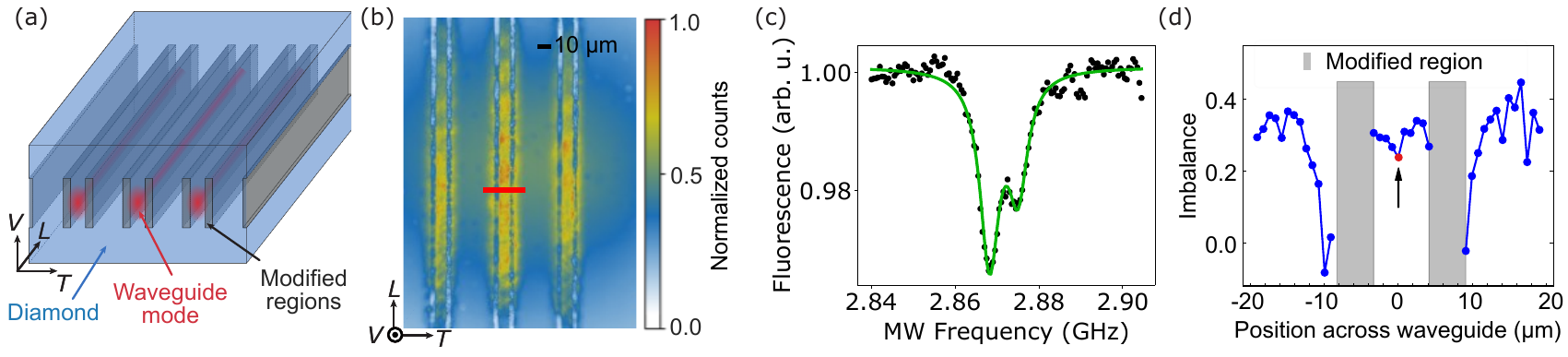}
\caption{{Experimental platform and asymmetric ODMR spectra of NV$^-$ centers in diamond in the waveguide.} (a) Schematic of a diamond sample with laser-modified regions and waveguides. The coordinate system defines the laboratory frame with the transverse ($T$), longitudinal ($L$), and vertical ($V$) directions. (b) Fluorescence map of NV$^-$ centers in the sample. The red line shows the path of the ODMR scan. (c) An exemplary ODMR spectrum. The black dots are experimental data; the green line is a fit using two Lorentzians. (d) Imbalance of the ODMR spectrum (dots) as a function of the position across the waveguide; the blue line is a guide to the eye. The red dot indicated by an arrow is the position where the ODMR spectrum of (c) is taken. }
\label{fig:ODMR_Imbalance}
\end{figure*}

The negatively charged nitrogen-vacancy (NV$^-$) center in diamond \cite{Doherty2013TheDiamond} has prospective applications ranging from biology to quantum information due to the long lifetime and coherence time of its spin states \cite{Astner2018Lifetime,Abobeih2018Coherence,Park2022Decoherence} and the fact that it can be initialized and read out optically at room temperature \cite{Schirhagl2014Nitrogen-VacancyBiology,Doherty2013TheDiamond,Suter2017Rev}. The electronic spin states of an NV$^-$ center are sensitive to changes in electric, magnetic, and strain fields, which promotes it as a potential quantum sensor \cite{Rondin2014RevMagnetometry,Barry2020RevMagnetometry,Zhang2023SingleNVRFsensor} with achievable atomic-scale resolution \cite{Sushkov2014NMRSingleProton,Abobeih2019AtomicScaleImaging}.
For sensing, single NV$^-$ centers can be incorporated into nanodiamonds or nanophotonic devices, allowing them to be used as atomic-scale probes for scanning probe microscopy or for in vivo imaging of cells or other biological specimens. 
As the measurement of the spin states depends on the collection of fluorescent light, the measurement sensitivity using only a single NV$^-$ center is relatively low, meaning that to achieve high sensitivity, the integration time of the measurement must be increased. Alternatively, using high-density ensembles of NV$^-$ centers enables faster measurements \cite{Taylor2008High-sensitivityResolution} and enhances sensitivity \cite{Hayashi2018OptimizationEnsemble}. However, the nanoscale resolution made possible by single NV$^-$ centers is lost. The recent development of laser-written waveguides in bulk diamond host crystal \cite{Courvoisier2016Waveguides,Sotillo2016} enables one to achieve relatively high photon collection efficiency \cite{Giakoumaki2022Quantum,Hoese2021IntegratedArrays,Hadden2018Integrated,Guo2024Laser}, constituting an important step towards using high-density NV$^-$ center ensembles for broadband integrated magnetic and electric field sensing \cite{Taylor2008High-sensitivityResolution,Acosta2009Diamonds}. To produce such structures, one uses femtosecond laser pulses to locally modify the diamond material by writing two parallel lines. Modification of the crystal structure and the induced strain modulate the refractive index, which confines an optical mode between the laser-modified lines \cite{Chen2011Laser-writing, Sotillo2018PolarizedDiamond}. On the other hand, strain affects the electronic states of NV$^-$ centers \cite{Davies1976NVstrain,Cai2014hybrid,Teissier2014StrainCoupling,Barfuss2019Spin-stressSystems,Udvarhelyi2018Spin-strainDiamond} and alters the collected signals. Thus, understanding the effects of strain and determining the strain field in processed samples is crucial for harnessing the full potential of the system in a wide range of quantum sensing applications.

There are various methods available to measure strain in the diamond crystal, such as using SiV centers \cite{Bates2021SiliconStrain}, the zero-phonon line of NV$^-$ centers \cite{Grazioso2013Measurement},
and quantum interferometry \cite{Marshall2022Highprecision}. The commonly used technique is optically detected magnetic resonance (ODMR) \cite{Oort1988ODMR}. Existing ODMR studies included determining the axial and non-axial strain using preferentially aligned NV$^-$ centers in unprocessed polycrystalline diamond \cite{Trusheim2016Wide} and in diamond nanocrystals hosting 1--4 NV$^-$ centers each \cite{Awadallah2023} or estimating the non-axial component based on single-NV$^-$ ODMR splitting \cite{Knauer2020Insitu}. However, to determine the entire strain tensor, one needs to go beyond the oversimplified equivalence between strain and electric field components \cite{Barfuss2019Spin-stressSystems}. Also, for practical use in high-density NV$^-$ ensembles, a method accounting for four possible center orientations in the host crystal is needed. Other studies \cite{Broadway2019Microscopic,Kehayias2019ImagingCenters} took this into account and used a static magnetic field to split the states and determine the strain components. However, most practical quantum sensing applications \cite{Hayashi2018OptimizationEnsemble,Shim2022Multiplexed,Sekiguchi2023Diamond} use experimental setups without an external magnetic field, and hence it is desirable to provide full strain tensor imaging under such conditions.

In this work, we present a method for extracting the relevant strain tensor components from continuous-wave zero-field ODMR signals. We first formulate a model of the randomly oriented ensemble of NV$^-$ centers in the presence of strain. Next, by fitting the ODMR spectra across a laser-written waveguide with the model results, we extract the strain profile across the waveguide structure. In this way, we demonstrate the potential of the zero-field ODMR as a strain characterization tool and determine the spatial profile of the relevant components of the strain tensor across a particular waveguide structure. We demonstrate the presence of a dominant compressive strain component in the direction transverse to the waveguide structure, accompanied by weaker vertical and shear strain components.

This paper is organized as follows. In Sec.~\ref{sec:experimental_methods}, we describe the experimental methods, including laser writing of waveguides and ODMR spectroscopy, and present the resulting ODMR spectra of NV$^-$ centers. Then, in Sec.~\ref{sec:theory}, we present our theoretical model. In Sec.~\ref{sec:strain_effects}, we discuss the effect of the strain on a single NV$^-$ center. In Sec.~\ref{sec:results}, we present the results of the fitting of the measured data with the modeling results and determine the strain profile across the waveguide. Finally, we conclude the paper in Sec.~\ref{sec:conclusions}. The appendix presents some additional information.

\section{Experiment}
\label{sec:experimental_methods}
\subsection{Laser-written waveguide devices}

Waveguides [Fig.~\hyperref[fig:ODMR_Imbalance]{\ref*{fig:ODMR_Imbalance}(a)}] are formed in the diamond crystal using a femtosecond laser writing technique with a custom setup based on a regeneratively amplified Yb:KGW system (Pharos, Light Conversion), previously described in Ref.~\cite{Sotillo2018PolarizedDiamond}. A Type II geometry is used where optical confinement is provided between two laser-written modification lines. The light confinement is caused by a reduction of the refractive index within the modification lines, where the focused femtosecond laser pulses cause amorphization and graphitization, and strain-induced increase of the refractive index between the two lines \cite{Sotillo2018PolarizedDiamond}. Laser-written diamond waveguides can be created in ultrapure CVD diamond with nitrogen concentrations less than 5 ppb \cite{Hadden2018Integrated,Hoese2021IntegratedArrays,Guo2024Laser}, as well as in diamonds with high nitrogen densities up to 200 ppm \cite{Guo2024Laser}.
In our experiment, we used a yellow diamond sample with a nitrogen content of approximately 200~ppm and native NV$^-$ concentration below 1~ppb, grown under high-pressure and high temperature (HPHT) conditions. In this sample, the laser-modified regions are created in the [100] direction at a depth of 25~\textmu{}m below the (001) plane, with a height of approximately 15~\textmu{}m, width approximately 5~\textmu{}m, separation between the two lines of 13~\textmu{}m (read off directly from a micrograph of a crystal facet displaying the cross-section of the structure), and length 3~mm. The waveguide is written with 50~mW average laser power. The laser writing process also produces a high density of vacancies in the diamond lattice within the modification lines \cite{Chen2019Laser,Griffiths2021Microscopic} that subsequently diffuse to the nearby regions. The sample was annealed at $1000^\circ\text{C}$ for 3~hours in a nitrogen atmosphere. Upon annealing, the vacancies become mobile and are ‘captured’ by substitutional nitrogen to form additional NV$^-$ centers.

Following annealing, we characterize the distribution of NV$^-$ centers over a broad area around the waveguides employing wide-field photoluminescence (PL).
These measurements allow efficient probing of the whole depth of the waveguide simultaneously.
Wide-field imaging is performed with a modified Nikon Ti-Eclipse inverted microscope equipped with a microwave loop.
A 532~nm fiber-coupled continuous-wave laser (CNI laser, MGL-III-532/50~mW) is used as an excitation source, delivering the power of $\sim30$~mW on the sample through a $40\times$ ($\text{NA} = 0.75$ and $\text{WD} = 0.66$~mm) refractive objective.
This setup provides illumination without preferential polarization, with a broad beam waist and small divergence, allowing the entire field of view to be simultaneously illuminated and imaged.
The induced PL is then collected through the same objective, the laser excitation is filtered out using a dichroic filter, and the fluorescence image is formed on a high-sensitivity CMOS camera (Hamamatsu ORCAFlash4.0 V2).

The results of the PL characterization are presented in Fig.~\hyperref[fig:ODMR_Imbalance]{\ref*{fig:ODMR_Imbalance}(b)}. One may observe the variation of intensity caused by the structurization superimposed with the Gaussian laser spot profile covering three waveguides across. The observed PL enhancement is consistent with forming more NV$^-$ centers in the structured area than in the pristine region. The estimated density of NV$^-$ centers in the waveguide after annealing is on the order of $10^{15}$~cm$^{-3}$ (10~ppb). The modified regions themselves remain dark on the PL map indicating low concentration of NV$^-$ centers, consistent with previous studies \cite{Sotillo2016}. 
Apart from intensity changes due to the laser spot profile, some inhomogeneity of PL intensity can be observed along the waveguides, indicating an inhomogeneous distribution of NV$^-$ centers in the structure. The variation in the PL along the waveguides is attributed to the nonuniform distribution of the background N
impurities in the HPHT crystal and to the submicrometer inhomogeneity
in the laser-formed sidewalls, which also contributes to the waveguide
scattering loss \cite{Sotillo2016}. 

\subsection{Wide-field ODMR scans of the waveguide}\label{sec:experiment-odmr}

We collected ODMR spectra in steps of 0.813~\textmu{}m across the waveguide, along a path marked with a red line in Fig.~\hyperref[fig:ODMR_Imbalance]{\ref*{fig:ODMR_Imbalance}(b)}, using the optical setup described above.
To this end, we applied a microwave (MW) field of $-20$~dBm with a frequency varying from 2.82~GHz to 2.92~GHz in discrete steps of 0.5~MHz.
For each frequency step, a PL image of the whole region was acquired, as described above. ODMR spectra are then extracted from the stack of PL images.
Our approach provides the PL/ODMR signal resolved in the plane to individual pixels of the CMOS camera.
To obtain a sufficient signal-to-noise ratio it is, however, necessary to aggregate signals from a certain region of interest. Based on the translational symmetry of the structure and our aim of providing a transverse profile, we decided to keep the region of interest narrow in the transverse direction (5~pixels, 0.81~\textmu{}m) and wider in the longitudinal one (120 pixels, 19.5~\textmu{}m).
While we illuminate a wide range of depths in the sample, the resolution of imaging in the vertical direction is ensured based on efficient collection only from a narrow range of depths around the focal plane, on the order of single \textmu{}m. For this study, we selected the focal plane depth using the waveguide as a reference, which means that our imaging mostly focuses on the upper part of the waveguide.
Overall, our method provides a micrometer resolution in the plane perpendicular to the waveguide.

The spectra are collected in a wider range of frequencies, while only a narrow range of 2.84--2.90~GHz contains the optical response of interest. The measured signal exhibits a tilted background and noise, manifested in particular at low and high frequencies, far from the physically meaningful features. The slope can be attributed to a frequency dependence of the overall luminescence intensity or to the characteristics of the detection system and is, therefore, irrelevant to the interpretation of the ODMR features. To address this, we preprocess the raw data by removing the slope. Then, we normalize each ODMR series so that they approximately span the 0.95--1 range of values.
All spectra are shown in the Supplementary Material \cite{supplement}, and the underlying raw and preprocessed data are provided as a publicly available dataset \cite{Alam2024Data}.
In Fig.~\hyperref[fig:ODMR_Imbalance]{\ref*{fig:ODMR_Imbalance}(c)}, we show an exemplary zero-field ODMR spectrum collected from a position around the center of the waveguide (marked as 0~\textmu{}m position; red point in Fig.~\hyperref[fig:ODMR_Imbalance]{\ref*{fig:ODMR_Imbalance}(d)}). Due to the high nitrogen concentration, the hyperfine structure is not energetically resolved \cite{Hayashi2018OptimizationEnsemble}. The observed spectrum shows a double-dip structure with pronounced asymmetry, which is typical for our collected data. 

Both of these features, splitting and asymmetry, have been observed in the past. However, it should be noted that their sources and experimental conditions were different from those discussed here.
In some studies, splitting of the ODMR spectrum into two dips was observed when probing ensembles of NV$^-$ centers, even for unprocessed samples without an external magnetic field. That splitting has a specific character, with a sharp spike-like feature at the frequency corresponding to the zero-field splitting $D$, dividing the expected single dip into two \cite{Zhu2014ObservationSystem}. The inner slopes are steeper than the outer ones and such a spectrum cannot be represented by two Lorentzians \cite{Mittiga2018ImagingDiamond,Matsuzaki2016OpticallyDiamond}. It was shown that this effect can be reproduced by averaging over an ensemble in which \mbox{$|\!\pm\!1\rangle$} states of individual NV$^-$ centers mix and split due to the presence of inhomogeneous broadening and fluctuations of magnetic and strain fields \cite{Matsuzaki2016OpticallyDiamond,Hayashi2018OptimizationEnsemble}. Given those fluctuations in the environment, the probability of \mbox{$|\pm1\rangle$} states being degenerate is found to be zero, which leads to split ensemble spectra. A similar effect is observed for ensembles subject to random electric fields \cite{Mittiga2018ImagingDiamond}.

In contrast to those cases, the spectrum observed here can be well represented as a sum of Lorentzian features (green line in Fig.~\hyperref[fig:ODMR_Imbalance]{\ref*{fig:ODMR_Imbalance}(c)}). Moreover, we expect that in our structure the probed ensemble is subject to a considerable and roughly uniform strain field. This situation allows us to treat the spectra as dominated by uniform fields with negligible impact of their fluctuations.

Asymmetric ODMR spectra were also observed in single-NV$^-$ center studies and the underlying imbalance of transition rates was described by invoking the effective fields containing the effect of local electric field fluctuations and strain \cite{Mittiga2018ImagingDiamond,Kolbl2019DeterminationSpins, Kumar2023high}. 
Here, we are dealing with a dense ensemble for which the influence of local charges should, at most, lead to the above-mentioned non-Lorentzian splitting of states, not observed by us, without introducing asymmetry \cite{Mittiga2018ImagingDiamond}. Thus, we expect that the asymmetry observed by us depends solely on the nature of the strain and its impact on the four differently oriented NV$^-$ species. 

To quantify the degree of the spectral asymmetry, we define the imbalance
\begin{align}\label{eq:imbalance_def}
\Tilde{I} = \frac{\Tilde{\alpha}_{-}-\Tilde{\alpha}_{+}}{\Tilde{\alpha}_{-}+\Tilde{\alpha}_{+}},
\end{align}
where $\Tilde{\alpha}_{-}$ and $\Tilde{\alpha}_{+}$ are the strengths of the dips estimated based on fitting the spectrum with two Lorentzians. 
Fig.~\hyperref[fig:ODMR_Imbalance]{\ref*{fig:ODMR_Imbalance}(d)} shows the imbalance across the waveguide with the laser-modified regions shaded in gray. The limits of these regions were determined a posteriori based on a clear, qualitative change of the ODMR spectra (see Appendix~\ref{appendix:OMDR-Modified-Region}), with an uncertainty of 1 or 2 data points and are consistent with the PL image in Fig.~\ref{fig:ODMR_Imbalance}(b) and with the directly determined geometry of the laser-modified regions. We observe that the ODMR spectra are consistently asymmetric. 

In the following, we will show that the split and asymmetric zero-field spectra can be consistently attributed to strain present in the laser-written waveguide and can be quantitatively reproduced by summing contributions from the four NV$^-$ species experiencing the same approximately uniform strain distribution differently and having different projections on the MW field. 

\section{Model}
\label{sec:theory}
In this Section, we present the theoretical model accounting for the strain that we then use to determine strain values based on experimental ODMR spectra. We first describe the strain effect on a single-NV$^-$ center and then account for the contribution from the four classes of differently oriented centers.

\begin{figure}[tb]
  \centering
\includegraphics[width=\linewidth]{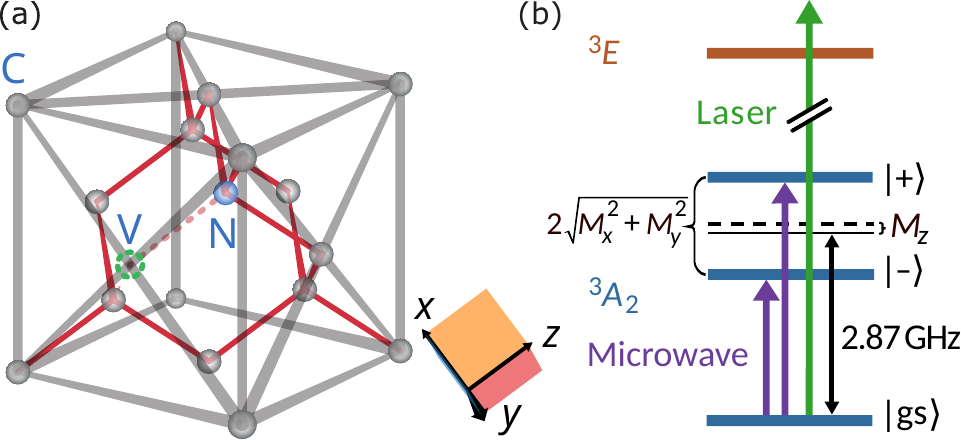}
	\caption{Structure and energy levels of an NV$^-$ center in diamond: (a) Schematic representation of an NV center in the diamond lattice. Carbon atoms are illustrated as gray spheres, with the nitrogen atom shown in blue and the lattice vacancy indicated by the green dashed line. Solid red lines represent the covalent bonds between the carbon atoms. The bond lines are drawn tapered to mimic the perspective. The inset shows the NV1 coordinates. (b) Energy level diagram of an NV$^-$ center. The ground state \( {}^{3}A_2 \) is a spin triplet (\( |\rm gs\rangle, |\pm\rangle \)), where (\( |\rm gs\rangle \)) is separated from $|\pm\rangle$ states by 2.87~GHz due to zero-field splitting. Optical excitation by a laser is shown in green, promoting electrons to the excited state (\( {}^3E \)). Microwave-induced transitions between the split ground states are shown in violet (singlet states, \( {}^1E \) and \( {}^1A_1 \) are not shown). In the presence of strain, the $|\pm\rangle$ states shift and split. $M_x$, $M_y$, and $M_z$ are the strain-amplitudes which are functions of strain components. }
  \label{fig:NV_crystal}
\end{figure}

The NV$^-$ center is a diamond defect created by replacing a carbon atom with a nitrogen atom adjacent to a carbon vacancy [Fig.~\hyperref[fig:NV_crystal]{\ref*{fig:NV_crystal}(a)}]. {For each NV$^-$ center, we use the reference frame with the $z$-axis in the direction connecting the nitrogen and the vacancy and the $x$ axis in one of the nitrogen-vacancy-carbon planes \cite{Udvarhelyi2018Spin-strainDiamond}.} In the following, lowercase coordinates refer to this NV$^-$-based reference frames. 

\begin{table}[tb!]
\caption{Spin-strain coupling strengths taken from Ref.~\cite{Udvarhelyi2018Spin-strainDiamond}.}
\begin{ruledtabular}
\begin{tabular}{cc@{\quad\quad}|cc}
\text{parameter} & \text{value (GHz)} & \text{parameter} & \text{value (GHz)} \\\hline
\( h_{43} \) & \( 2.3\pm0.2 \) & \( h_{26} \) & \( -2.83\pm0.07 \) \\
\( h_{41} \) & \( -6.42\pm0.09 \) & \( h_{15} \) & \( 5.7\pm0.2 \) \\
\( h_{25} \) & \( -2.6\pm0.08 \) & \( h_{16} \) & \( 19.66\pm0.09 \)
\end{tabular}
\end{ruledtabular}
\label{tab:spin-strain-coupling-constants}
\end{table}

The ground state of an NV$^-$ center is a spin triplet ($|0\rangle$, \mbox{$|\!\pm\!1\rangle$}). The Hamiltonian for the orbital ground state of an NV$^-$ center in the absence of a static magnetic field and in the presence of strain and MW driving can be written as \cite{Udvarhelyi2018Spin-strainDiamond}
\begin{subequations}
\label{eqn:full_Hamiltonian}
\begin{align}
   H =&\ hDS^2_z + \lambda (t)\widehat{\bm{B}}_{\mathrm{MW}} \cdot \bm{S} + H_{\varepsilon}\label{eqn:full_Hamiltonian_D},  \\
   H_{\varepsilon}/h =&\ M_{z}S^2_z+M_{x}(S_y^2-S_x^2)+M_{y}\{S_x,S_y\} \nonumber\\
    & +N_x\{S_x,S_z\}+N_y\{S_y,S_z\}\label{eqn:full_Hamiltonian_MN},
\end{align}    
\end{subequations}
where $h$ is Planck's constant, $D=2.87$~GHz is the zero-field splitting between $|0\rangle$ and \mbox{$|\!\pm\!1\rangle$} states, $\widehat{\bm{B}}_{\mathrm{MW}}$ is the unit vector along the MW magnetic field driving the transitions between $|0\rangle $ and \mbox{$|\!\pm\!1\rangle$} which is used to manipulate the population of the spin states, $\lambda (t)$ is proportional to the MW field amplitude, $S_j$ are the dimensionless spin-1 matrices and $M_{j}$, $N_{j}$ ($j=x,y,z$) are the strain amplitudes,
\begin{subequations}
\label{eqn:strain_amplitudes}
\begin{align}
  M_{z} & = \left[h_{41} (\varepsilon_{xx} + \varepsilon_{yy}) + h_{43} \varepsilon_{zz}\right], \\
  M_{x} & = \frac{1}{2} \left[ h_{16}  \varepsilon_{xz} - \frac{1}{2} h_{15} (\varepsilon_{xx} - \varepsilon_{yy}) \right], \\
  M_{y} & = \frac{1}{2} \left( h_{16} \varepsilon_{yz} + h_{15} \varepsilon_{xy} \right), \\
N_x &= \frac{1}{2} \left[ h_{26}  \varepsilon_{xz} - \frac{1}{2} h_{25} (\varepsilon_{xx} - \varepsilon_{yy}) \right], \\
N_y & = \frac{1}{2} \left(h_{26}  \varepsilon_{yz}+h_{25}\varepsilon_{xy}\right),
  \end{align}    
\end{subequations}
where $\varepsilon_{ij}$ are strain tensor elements in the NV$^-$ frame, while $h_{ij}$ are the spin-strain coupling strengths taken from Ref.~\cite{Udvarhelyi2018Spin-strainDiamond} and given in Tab.~\ref{tab:spin-strain-coupling-constants}.

Due to the tetrahedral structure of the diamond crystal, for a fixed vacancy position, there are four possible positions for an adjacent nitrogen atom to form an NV$^-$ center. Thus, there are four possible orientations of NV$^-$ centers in the crystal lattice. 
All expressions for the four cases are formally identical when expressed in the local NV$^-$ coordinates. However, due to the different orientations of NV$^-$ centers, the coordinates of the MW field and strain in the NV$^-$-based coordinate frames are different, resulting in distinct eigenvalues and transition rates.

We define the global coordinate frame through the geometry of the waveguide. We refer to the waveguide direction as longitudinal ($L$), the other in-plane direction as transverse ($T$), and the out-of-plane one as vertical ($V$) [Fig.~\hyperref[fig:ODMR_Imbalance]{\ref*{fig:ODMR_Imbalance}(a)}]. 
The translational symmetry of the structure in the $L$ direction imposes $\varepsilon_{LL}=\varepsilon_{LV}=\varepsilon_{LT} = 0$, and hence reduces the number of non-zero elements to three. Thus, the strain is characterized by its transverse and vertical components, as well as the corresponding shear component, and can be written as
\begin{equation}
\varepsilon_{\mathrm{lab}} = \begin{pmatrix}\varepsilon_{TT} & 0 & \varepsilon_{TV}\\ 0 & 0 & 0\\ \varepsilon_{TV} & 0 & \varepsilon_{VV}\end{pmatrix}.
\end{equation}
We must stress that this symmetry-based simplification is actually a slight approximation given some unavoidable inhomogeneity of the modification lines in the longitudinal direction. However, it is certainly not as pronounced as the PL inhomogeneity observed in Fig.~\hyperref[fig:ODMR_Imbalance]{\ref*{fig:ODMR_Imbalance}(a)}, as the latter is mainly due to variation in NV$^-$ concentration. 

For a particular NV$^-$ center, the strain is transformed to the NV$^-$ reference frame (see Appendix~\ref{appendix:transformations}), and the Hamiltonian [Eq.~\eqref{eqn:full_Hamiltonian}] is then diagonalized. The spectrum for the ground-state manifold of the NV$^-$ center, shown in Fig.~\hyperref[fig:NV_crystal]{\ref*{fig:NV_crystal}(b)}, is composed of the ground state $|\mathrm{gs}\rangle$, with the predominant contribution from the $|0\rangle$ state, and a pair of excited states $|\pm\rangle$, mostly built of the basis states \mbox{$|\!\pm\!1\rangle$}. The corresponding energies are $E_{\mathrm{gs}}$ and $E_{\pm}$. The relevant MW-driven transitions are those between the state $|\mathrm{gs}\rangle$ and the states $|\pm\rangle$. For these two transitions, we determine the rates that directly translate into light intensities, 
\begin{equation}\label{eq:rates}
\alpha_{\pm} = \left| M_{\pm} \right|^2 = \left| \left\langle 
\mathrm{gs} \left| \widehat{\bm{B}}_{\mathrm{MW}} \cdot \bm{S} \right| \pm
\right\rangle\right|^2,
\end{equation}
as well as their spectral positions $h\nu_\pm=E_\pm - E_{\mathrm{gs}}$. Each of these transitions leads to depletion of the ground state and, hence, reduces the optical fluorescence, which is manifested by a dip in the ODMR spectrum. We model these dips by Lorentzians with a certain width $\gamma$, 
\begin{equation*}
    L(\nu) =  \frac{\gamma}{\nu^2+\gamma^2},
\end{equation*}
where \(\nu \) is the frequency, from which we construct the spectrum of a single NV$^-$ center,
\begin{equation}
    f(\nu) = 1 - A\left[\alpha_{-} L\left(\nu-\nu_{-}\right)+\alpha_{+} L\left(\nu-\nu_{+}\right)\right],
    \label{eqn:Lorentzian_fit}
\end{equation}
with an arbitrary scaling amplitude $A$.

In the case of an ensemble, we assume that the density of the NV$^-$ centers is high enough compared to the range of variation of the strains so that a large number of them contribute from a volume in which the strain can be considered locally uniform. This is ensured by our experimental conditions outlined in Sec.~\ref{sec:experiment-odmr} with a micrometer resolution in the plane perpendicular to the waveguide and a lower one in the longitudinal direction, for which we expect no strain variation. Therefore, we assume that each orientation is equally probable. Additionally, all NV$^-$ species are tilted at the same angle to the optical $z$ axis and thus to the low-divergence unpolarized excitation beam, so that the total signal is composed of equal contributions from NV\(^-\) centers with all four different orientations. 
To calculate the ensemble ODMR spectrum, the strain is first transformed to the crystallographic frame (accounting for the orientation of the laser-written device in the crystal) and then rotated to the four NV\(^-\) frames (see Appendix~\ref{appendix:transformations}). Then the spectrum for each orientation is determined by diagonalizing the Hamiltonian and determining the transition strengths. The simulated ODMR spectrum is then a sum,
\begin{equation}
    F(\nu) = \sum_{i=1}^{4}f_{i}(\nu),
    \label{eqn:Lorentzian_fit_all}
\end{equation}
where \( f_{i}(\nu) \) is the single-NV\(^-\) spectrum given by Eq.~\eqref{eqn:Lorentzian_fit}, evaluated for the \( i^{\text{th}} \) orientation of the NV\(^-\) center.

\section{ODMR spectra and strain effects}
\label{sec:strain_effects}

In this section, we present an overview of the properties of the model and the role of various 
components of the strain. To understand the impact of strain on the response of an NV$^-$ ensemble, we first analyze a simpler case of a single center. The terms in Eq.~\eqref{eqn:full_Hamiltonian_MN} proportional to $N_x$ and $N_y$ induce only couplings between the $|0\rangle $ state and the \mbox{$|\!\pm\!1\rangle$} states, which are small compared to the zero-field splitting between these states induced by the first term in Eq.~\eqref{eqn:full_Hamiltonian_D}. 
Therefore, these terms give rise to small corrections in the 2nd order of the perturbation theory and can be neglected in the present general discussion, even though their magnitudes may be comparable to the $M_{x,y}$ terms.

\subsection{MW-driven transitions}
\label{sec:transitions}

Neglecting the terms proportional to $N_x$ and $N_y$, one can conveniently analytically diagonalize Eq.~\eqref{eqn:full_Hamiltonian} with $\lambda (t)= 0$ \cite{Mittiga2018ImagingDiamond}. This yields the energies
\begin{equation}
    E_{\mathrm{gs}} = 0,\quad
    E_{\pm} = \left( D+M_{z}\pm \sqrt{M_{x}^2+M_{y}^2}\right),
    \label{eqn:eigenenergy}
\end{equation}
as well as the corresponding eigenstates
\begin{subequations}
\label{eqn:eigenbasis}
\begin{align}
|\mathrm{gs} \rangle &= |0\rangle,\\
|\pm\rangle &= \frac{1}{\sqrt{2}}\left(\pm e^{i\phi_{\mathrm{str}}}|-1\rangle+|1\rangle \right),
\end{align}    
\end{subequations}
where
$\phi_{\rm str} = \tan^{-1}(M_{y}/M_{x})$ is the effective strain ``vector'' in-plane angle. 
We see that strain at the position of the NV$^-$ center leads to the mixing of \mbox{$|\!\pm\!1\rangle$} eigenstates into their split superpositions $|\pm\rangle$ [Fig.~\hyperref[fig:NV_crystal]{\ref*{fig:NV_crystal}(b)}]. The Hamiltonian written in the eigenbasis reads,
\begin{align}\label{HNV_Bz_0_eigenbasis}
    H =& E_{-}|-\rangle\langle -|+E_{+}|+\rangle\langle +| \notag\\
        &+[ h\lambda'(t)M_{-} |0\rangle\langle -| +h\lambda'(t)M_{+} |0\rangle\langle +|+ {\rm H.c.}],
\end{align}
where
$\lambda'(t) = \frac{2}{3} \lambda(t)$.

Using this result, we find the transition rates from Eq.~\eqref{eq:rates}. Up to normalization, these can be written as
\begin{equation}
\alpha_{\pm} = \frac{1\pm\cos\left(2\phi_{\mathrm{MW}}+\phi_{\mathrm{str}}\right)}{2},
\label{eqn:transition_probability_analytical}
\end{equation}
where $\phi_{\rm MW}$ is the azimuthal angle between the NV$^-$-plane MW field component and the $x$-axis of an NV$^-$ center.

\subsection{Effect of strain on a single \texorpdfstring{NV$^{-}$}{} center}

\begin{figure*}[tb]
    \includegraphics{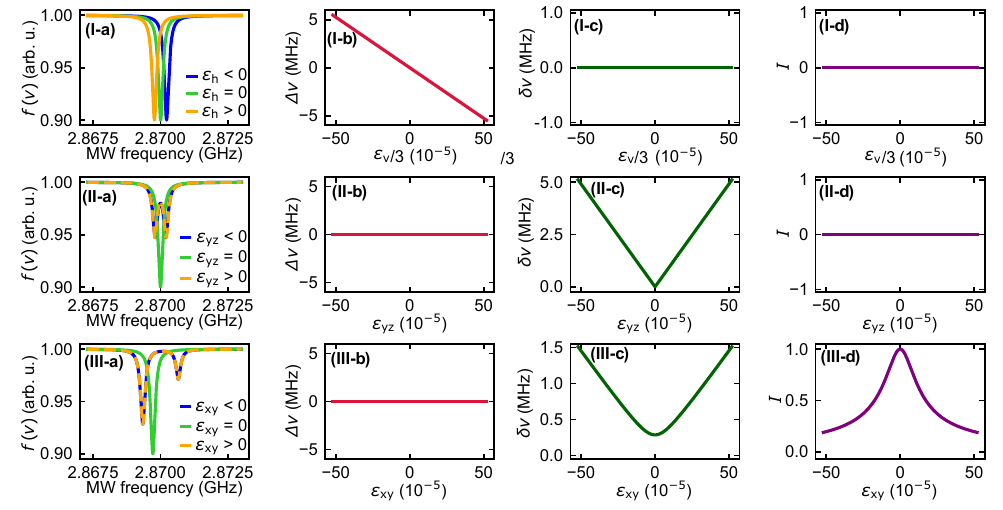}
    \caption{Effects of strain on a single NV$^-$ center: (a) ODMR spectrum for three strain values (blue line: positive strain, orange line: negative strain, green line: no strain); (b) Spectral shift $\Delta \nu$ of the ODMR dips from the position corresponding to the zero-field splitting ($D$); (c) Splitting $\delta \nu$ of the $|\pm\rangle$ states; (d) Imbalance $I$ of the ODMR dips. We studied three different types of strain: (I) Volumetric strain, (II) Shear strain in the $yz$ plane in the absence of axial strains, and (III) Shear strain in the $xy$ plane in the presence of strain components $\varepsilon_{ xx} =10\times10^{-5}$ and $\varepsilon_{ yy} = -10\times10^{-5}$ with $\varepsilon_{ zz} = 0$.}
    \label{fig:strain_effect}
\end{figure*}

The two states $|\pm\rangle$ enable two MW-active transitions that give rise to two lines in the ODMR spectrum at frequencies $\nu_\pm$.
In this section, we discuss how the presence of strain in the crystal affects the properties of NV$^-$ centers. We focus on a single center and analyze the impact of strain on the ODMR spectrum in terms of three characteristic features: the shift of the spectral lines, the splitting between them, and the asymmetry (imbalance) between them. We use the approximate model with neglected $N_{x,y}$ terms, hence $E_{\mathrm{gs}}=0$ and the positions of the lines directly reflect the spectrum of the NV$^-$ center, $h\nu_\pm = E_\pm$. 
 
The strain can change the overall position of the ODMR spectral lines, adding to the zero-field splitting of the NV$^-$ center $D$. The shift can be quantified by the deviation of the mean energy of the corresponding eigenstates [Eq.~\eqref{eqn:eigenenergy}] from their strain-free zero-field position,
\begin{equation}
\label{eqn:shifting}
\Delta \nu = (\nu_{+}+\nu_{-})/2 -D= M_z.
\end{equation}
It depends on $M_z$, which is linear in axial strain components. Hence, these components are responsible for the overall shift of the ODMR lines. 

The strain can also induce the splitting of the states $|\pm \rangle$, which is expressed as
\begin{equation}
\label{eqn:splitting}
\delta \nu = (\nu_{+}-\nu_{-}) = 2\sqrt{M_x^2+M_y^2},
\end{equation}
and depends on $M_x$ and $M_y$, which are functions of both axial and shear strain components. However, note that only a difference of axial components is involved, which translates to pure shear strain in a 45~deg rotated coordinate frame. Thus, this effect, in fact, depends only on shear strain. 

The theoretically predicted imbalance $I$ is defined in analogy to Eq.~\eqref{eq:imbalance_def}
\begin{equation*}
    I = \frac{\alpha_{-}-\alpha_{+}}{\alpha_{-}+\alpha_{+}}
    = -\cos(2\phi_{\mathrm{MW}}+\phi_{\mathrm{str}}).
\end{equation*}
The imbalance is caused by nonzero $M_x$ or $M_y$ that enter $\phi_{\mathrm{str}}$, but in detail it depends on the in-plane MW orientation angle $\phi_{\rm MW}$. 

The theoretically predicted effect of strain on these three characteristics of the ODMR spectrum is presented in Fig.~\ref{fig:strain_effect}. In the subsequent rows of this figure, we study the effect of three classes of strains. For each of these, we first plot spectra for selected values of the strain amplitudes in the leftmost column. The ODMR signal is plotted using Eq.~\eqref{eqn:Lorentzian_fit}, where the strain enters the ODMR signal through the eigenenergies [Eq.~\eqref{eqn:eigenenergy}] and transition rates [Eq.~\eqref{eqn:transition_probability_analytical}], which are the parameters of the double Lorentzian function. In each panel, the blue, green, and orange lines represent the cases of positive, zero, and negative strains, respectively. The dependence of the three characteristic quantities (shift, splitting, and imbalance) on the magnitudes of the relevant strain components is then plotted in the other three columns. The results presented are obtained from the approximate model ($N_x=N_y=0$) for $\phi_{\mathrm{MW}}=\pi/2$, i.e., with the MW polarized along the $y$ axis. 

To illustrate the role of various strain components, we study the effect of three particular types of strain.

I. \textit{Volumetric strain} (first row in Fig.~\ref{fig:strain_effect}): Fig.~\hyperref[fig:strain_effect]{\ref*{fig:strain_effect}(I-a)} illustrates the effect of volumetric strain, $\varepsilon_{\mathrm{v}} = \text{Tr}\varepsilon$, on the ODMR lines. Compressive (negative) strain shifts the ODMR features towards higher energy (blue line), while it is the opposite for tensile strain (orange line). Fig.~\hyperref[fig:strain_effect]{\ref*{fig:strain_effect}(I-b)} demonstrates the linear variation of the shift. Fig.~\hyperref[fig:strain_effect]{\ref*{fig:strain_effect}(I-c)} shows no spectral splitting, as there is no shear strain and the in-plane axial strains are equal, hence no coupling between the excited states. Obviously, no imbalance can be found as the two transitions are degenerate.

II. \textit{Pure shear strain in the $yz$ plane} (second row in Fig.~\ref{fig:strain_effect}): Here, we assume $\varepsilon_{yz}\neq 0$ while all other components are 0. For this type of strain, the states $|\pm\rangle$ split, which is manifested by the splitting of the ODMR lines in Fig.~\hyperref[fig:strain_effect]{\ref*{fig:strain_effect}(II-a)}). We plot the ODMR spectra for negative (blue line) and positive (dashed orange line) values of $\varepsilon_{yz}$, and in the absence of strain (green line). Since the magnetooptical response depends on the magnitude of the shear strain, the spectra for positive and negative $\varepsilon_{yz}$ are identical. We do not observe any shift of the lines since there are no axial strain components [Fig.~\hyperref[fig:strain_effect]{\ref*{fig:strain_effect}(II-b)}]. The line splitting varies linearly with strain [Fig.~\hyperref[fig:strain_effect]{\ref*{fig:strain_effect}(II-c)}]. Again, we do not observe any imbalance in the spectrum [Fig.~\hyperref[fig:strain_effect]{\ref*{fig:strain_effect}(II-d)}]. The lack of asymmetry is due to the interplay of the MW orientation and the effective strain ``vector'' angle (of $\pi/2$ given $M_x=0$ and $M_y\neq0$). The two transitions couple to a magnetic field polarized along diagonal directions ($\hat{x}+\hat{y}$ and $\hat{x}-\hat{y}$), so for $y$-polarized MW chosen here they both couple with the same strength. Note that by varying the MW angle, we would get a full oscillation of imbalance as $I=-\sin(2\phi_{\mathrm{MW}})$. 

III. \textit{Shear strain in the $xy$ plane on top of another in-plane shear strain contribution}: Here, we apply varying $\varepsilon_{xy} \neq 0$, with fixed $\varepsilon_{xx} = - \varepsilon_{yy} = 10^{-4}$, while $\varepsilon_{zz} =  \varepsilon_{yz} = \varepsilon_{xz}=0$ . In this case, the $|\pm\rangle$ states mix and split, and the signal additionally exhibits an imbalance [Fig.~\hyperref[fig:strain_effect]{\ref*{fig:strain_effect}(III-a--III-d)}]. The nonzero in-plane diagonal strain elements of opposite signs correspond to pure shear strain (in a 45-deg-rotated basis) and result in transitions polarized along $\hat{x}$ and $\hat{y}$ so that one of them is fully colinear with the applied MW, and the other is dark. When another shear strain element is introduced ($\varepsilon_{xy}$), the polarizations of transitions get rotated, resulting in varying and unequal brightness of the two. However, we do not observe any shift of the spectral lines since the in-plane axial components have opposite values [Fig.~\hyperref[fig:strain_effect]{\ref*{fig:strain_effect}(III-b)}]. Lastly, Fig.~\hyperref[fig:strain_effect]{\ref*{fig:strain_effect}(III-d)} demonstrates that we can adjust the imbalance by tuning the shear strain component.

To provide a general understanding, we may refer to the symmetry arguments. Originally, in the presence of three-fold rotational symmetry, the $\ket{\pm 1}$ states are degenerate as they transform according to the same irreducible representation of the $C_{3v}$ point group~\cite{Dresselhaus2008}. They are split from the $\ket{0}$ state due to the lack of full tetrahedral symmetry (the $z$ axis of the NV$^-$ center is distinguished). Any strain that conserves the $C_{3v}$ symmetry can only lead to further splitting between $\ket{0}$ and $\ket{\pm 1}$, and this is what we observe for volumetric strain. It would also be the case for axial strains as long as the in-plane components are equal and hence preserve symmetry. Any shear strain, also hidden in the form of a difference of axial components in the $xy$ plane, breaks the rotational symmetry and, in principle, causes the appearance of coupling between $\ket{\pm 1}$ states, leading to their mixing and splitting. The states become mixed into combinations that, depending on the mixing phase, may turn out to be (partially) bright and dark with respect to a MW with a specific in-plane component. This effect is reflected in the dependence of transition rates on the interplay of the in-plane MW and effective strain ``vector'' angles in Eqs.~\eqref{eqn:transition_probability_analytical}. Finally, if we first split the states with one strain component and then try to drag them through resonance with another strain component, the levels anticross, as expected in the situation where all the symmetries have been broken.

\section{ODMR-based strain characterization}
\label{sec:results}

\begin{figure}[tb]
    \includegraphics{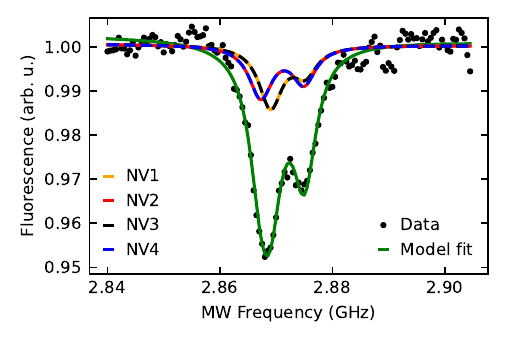}
    \caption{Fitting the double-dip, asymmetric ensemble ODMR spectrum data (points) with our model at position 2.49~\textmu{}m. We extract the strain values from the fitting (green line): {$\varepsilon_{TT} = -103.9\times10^{-5}$, $\varepsilon_{VV} = -27.1\times10^{-5}$, $\varepsilon_{TV} = \pm8.4\times10^{-5}$}, and $D = 2.86670\pm0.00014$~GHz (the uncertainty represents the fitting uncertainty as given by the least squares procedure). With these strains, we simulate the individual NV$^-$ ODMR spectra (orange, red, black, and blue lines), observing that strain enters differently for different NV$^-$s, resulting in the overall spectrum having two dips and being asymmetric.}
    \label{fig:ODMR_experimental_theory_fit}
\end{figure}

In this section, we fit the ODMR spectra collected at a sequence of points across the waveguide structure with the theory presented in Sec.~\ref{sec:theory} (using the full model including the terms $N_{x,y}$) and extract the components of the strain tensor across the laser-written waveguide. In the fits, we additionally allow a linear component to remove a residual tilt in the data after removing the tilted background, as discussed above. Fitting the data integrated over a certain area assumes reasonable uniformity of the strain within this area, which may be satisfied only approximately. In fact we characterize strain averaged over possible small-scale fluctuations. On the other hand, our results show that this locally averaged strain relaxes in space on a length scale that is resolved by our experiment. We set the ODMR linewidth to $\gamma=2.5$~MHz, which minimizes the overall least-squares residuals over the full set of fitted data series. This linewidth is consistent with the inhomogeneous dephasing time $T_2^*$ measured on the same sample \cite{Guo2024Laser}. We allow the zero-field splitting to be a free fitting parameter, even though, in principle, it is a material constant. This choice consistently improves the fitting results and can be justified in two ways. First, the translational symmetry is only an approximation given some unavoidable inhomogeneity of the laser-modified regions. A correction to the value of $D$ allows us to account for a residual volumetric strain that may appear due to this inhomogeneity. Second, the material parameters may vary slightly due to high doping and laser writing. As we show below, the obtained variation of $D$ is very small and remarkably systematic across the waveguide. The value of the parameter $D$ obtained in this way from the fitting will be referred to as the \textit{effective zero-field splitting}.

An exemplary fit to an ODMR spectrum measured inside the waveguide [the same as in Fig.~\hyperref[fig:ODMR_Imbalance]{\ref*{fig:ODMR_Imbalance}(c)}] is shown in Fig.~\ref{fig:ODMR_experimental_theory_fit} (green line). From the fitting, we extract the strain components, \( \varepsilon_{TT} = -103.9 \times 10^{-5} \), \( \varepsilon_{VV} = -27.1 \times 10^{-5} \), \( \varepsilon_{TV} = \pm 8.4 \times 10^{-5} \) (the results are insensitive to the sign of shear strain), and $D = 2.86670\pm0.00014$~GHz. Repeating the fitting with $N_{x,y}$ set to zero yields values that differ from the original ones relatively by at most $10^{-3}$, which confirms the validity of the simplified model used in the discussion of Sec.~\ref{sec:strain_effects}. In Fig.~\ref{fig:ODMR_experimental_theory_fit}, we also show the decomposition of the total ODMR signal into the contributions from the four NV$^-$ orientations. We find that they divide into two pairs that shift and split differently, as could be expected on the basis of symmetry considerations.

\begin{figure}[tb]
    \includegraphics{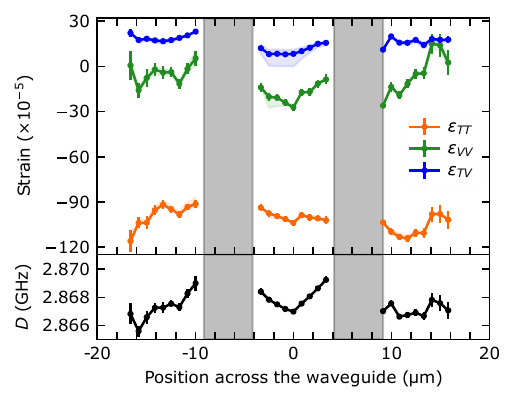}
    \caption{Fitting results. Upper: Strain profiles across the waveguide. The orange curve shows the transverse strain component \( \varepsilon_{TT} \), which is compressive. The shaded region represents the uncertainty due to the Lorentzian dip width ($\gamma$). The green curve shows the vertical strain component \( \varepsilon_{VV} \), which is also compressive. The blue curve shows the shear strain component \( \varepsilon_{TV} \), which is insensitive to its sign; therefore, we plotted its absolute value. Shear strain in the waveguide region shows a strong dependence on $\gamma$. Lower: the effective zero-field splitting of a single NV$^-$ center slightly deviates from 2.870~GHz across the waveguide. This variation is systematic inside the waveguide.}
    \label{fig:strains}
\end{figure}
 
In the same way, we fit the model to ODMR spectra collected from different positions across the waveguide, along the path shown as the red line in Fig.~\hyperref[fig:ODMR_Imbalance]{\ref*{fig:ODMR_Imbalance}(b)}, and extract the three strain components expressed in the laboratory frame and the effective zero-field splitting parameter. All the fits, as well as the initial and final values of the fitting parameters, are given in the Supplementary Material \cite{supplement}. The upper panel in Fig.~\ref{fig:strains} shows the three strain components extracted from the fitting and plotted across the waveguide. We exclude the laser-modified regions where the signal is weak because of a low density of NV$^-$ centers, and our model does not apply because of a different material structure and thus qualitatively different magneto-optical response (see Appendix~\ref{appendix:OMDR-Modified-Region}). The shaded regions show the range of the fitting values obtained with the assumed linewidths $\gamma$ ranging from 2.1 to 2.9~MHz, showing weak dependence of the fitting results on this parameter. The solid line corresponds to \(\gamma = 2.5\)~MHz, for which the sum of residuals of the points inside the waveguide region is minimal.
The strain within the waveguide is dominated by the transverse strain component (orange curve, \(\varepsilon_{TT}\)), which is compressive both inside and outside of the waveguide. The green curve represents the vertical strain component (\(\varepsilon_{VV}\)) which is also compressive. The blue curve represents the shear strain component (\(|\varepsilon_{TV}|\)), for which we show absolute values, as ODMR spectra are insensitive to its sign. This strain component inside the waveguide shows larger uncertainty related to the selected value of \(\gamma\). We also observe a small variation of the effective zero-field splitting parameter \(D\) across the waveguide, as shown in the lower panel in Fig.~\ref{fig:strains}. The relative magnitude of this variation is on the order of $10^{-3}$. Moreover, within the waveguide, this variation is remarkably systematic, with an increase towards the modified regions, suggesting a real physical effect behind this parameter change.

Outside the waveguide, the fitting results show more fluctuations and some asymmetry between the two sides of the structure. 
This may indicate that the results are more uncertain than indicated by the error bars in Fig.~\ref{fig:strains}. Indeed, the uncertainty shown here by the error bars is just the technical error of the fitting. The uncertainty of the obtained results is much higher due to the weak signal outside of the structure and can be estimated by comparing the results to the left and to the right of the waveguide structure, which should be symmetric in principle. Note that this imperfect symmetry is consistent with the inhomogeneity of the system manifested in the PL map in Fig.~\ref{fig:ODMR_Imbalance}(b). 

Our results are comparable to the previously reported stress tensor components studied using polarized micro-Raman spectroscopy \cite{Sotillo2018PolarizedDiamond} (see Appendix~\ref{appendix:stress}). In both cases, the transverse stress component at the center of the waveguide is compressive with values in the range of 1.05--1.15~GPa. The two structures differ in their crystallographic orientations, defect densities, manufacturing details, and signal collection volumes, which precludes direct comparison of the results. In addition, our model includes shear strain or stress components, which was not considered in that earlier work.

\section{Conclusions}
\label{sec:conclusions}
We have fabricated linear waveguide structures in diamond with a high density of NV$^-$ centers and performed a systematic characterization with wide-field ODMR at different positions across a waveguide. We have found the zero-field ODMR spectra to be consistently split and asymmetric. The Lorentzian character of collected signals suggests that the observed features are not due to averaging over an ensemble subject to an inhomogeneous environment but rather due to a strain field close to uniform in the probed volume. This allowed us to formulate a model for the optical response of an MW-driven ensemble of differently oriented NV$^-$ centers in which individual strain tensor components can be treated as fitting parameters.

To gain a general qualitative understanding of the effects of strain on ODMR spectra, we first simulated a single NV$^-$ center with a fixed MW magnetic field orientation and demonstrated that the asymmetry of the spectra can be controlled by adjusting specific strain values and interpreted in terms of rotational symmetry breaking. Next, we utilized the model and the collected ODMR signals to determine the strain profile induced in the crystal during the laser writing of the waveguide structure. We found a dominant compressive axial strain in the direction transverse to the waveguide structure and weaker vertical and shear strain components. Thus, we have proposed a method for the full spatial characterization of the strain tensor in highly symmetric waveguides that does not require using a magnetic field. While we have demonstrated its operation by mapping strain across a specific structure, the method is applicable to other systems.

Our findings show that NV$^-$-rich laser-written waveguide structures that provide strong optical response can be fully characterized in terms of built-in strain using the commonly used technique of zero-field ODMR. Such feedback should allow for feasible quality control of created structures. It should also pave the way for their use for measuring externally applied strain and fields by model-assisted examination of the variation in the ODMR spectra, for which precise determination of the built-in deformations and their effects is necessary.

Overall, our study contributes insights into the behavior of NV$^-$ centers in waveguide structures. However, it is essential to be aware of its limitations. Our method requires all four NV$^-$ orientations to be uniformly represented in the sampled volume and the waveguide to obey translational symmetry. We did not account for the effects of random electric and magnetic fields, which is justified, however, by the dominant role of the probed structural strain on the modeled spectra. Nonetheless, further work to account for some of those effects may be valuable. It is also crucial to investigate how far one can go in increasing the spatial resolution of this method, which means reducing the size of the sampled ensemble.

\section*{DATA AVAILABILITY}
The raw data, as well as the pre-processed data used in the fitting, are provided as a publicly available dataset \cite{Alam2024Data}.

\begin{acknowledgments}
We thank Belen Sotillo and Akhil Kuriakose for fruitful discussions. This project has received funding from the European Union’s Horizon 2020 research and innovation programme under the Marie Sk{\l}odowska-Curie ITN project LasIonDef (GA No.~956387). MG acknowledges support from the National Science Centre (Poland) under Grant No. 2015/18/E/ST3/00583. DW acknowledges financial support by the Science Foundation Ireland (SFI) under grants nos. 18/RP/16190 and 22/PATH-S/10656. AJB and JPH acknowledge the financial support provided by EPSRC via Grant No. EP/T017813/1 and EP/03982X/1. VB acknowledges the support of the Alexander von Humboldt Foundation. AB gratefully acknowledges the financial contribution from MAECI, "Italy-Israel joint programme --- 2023 scientific track" within the project PRECIOUSMRI. SME is thankful for the support from the projects QuantDia (FISR2019-05178) and PNRR PE0000023 NQSTI funded by MUR (Ministero dell'Universit\`{a} e della Ricerca). MG, DW, and PM acknowledge support from the Alexander von Humboldt Foundation in the framework of the Research Group Linkage Programme funded by the German Federal Ministry of Education and Research.

\end{acknowledgments}

\begin{appendix}

\section{\MakeUppercase{ODMR spectra in the laser-modified region}}
\label{appendix:OMDR-Modified-Region}

\begin{figure}[tb]
    \centering
    \includegraphics[width=\linewidth]{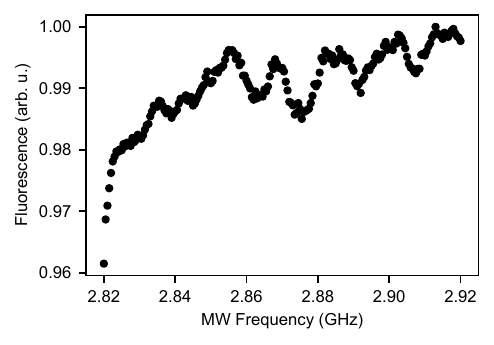}
    \caption{An exemplary ODMR spectrum collected from the laser-modified region at the position of 8.31~\textmu{}m in terms of Fig.~\hyperref[fig:ODMR_Imbalance]{\ref*{fig:ODMR_Imbalance}(d)}.}
    \label{fig:ODMR-Modified-Region}
\end{figure}

As mentioned in the main text, the ODMR spectra collected from the laser-modified regions are qualitatively different from those from the unprocessed material regions due to the difference in material structure, which allows us to determine the range of these regions based on the ODMR data. Fig.~\ref{fig:ODMR-Modified-Region} shows an example of such a spectrum, taken at the position of approximately 8~\textmu{}m. This spectrum qualitatively differs from those taken in the other areas (Fig.~\ref{fig:ODMR_Imbalance}(c)), showing multiple features beyond the standard magneto-optical response of an NV$^-$-rich diamond sample.

\section{\MakeUppercase{Transformation to NV reference frames}}
\label{appendix:transformations}

The strain and the MW field are parametrized by their components in the ``laboratory frame'' $(T,L,V)$ bound to the laser-written waveguide structure. In order to simulate the ODMR response, they need to be transformed to the NV reference frame for each of the four NV orientations. This is done in two steps.

In the first step, we account for the orientation of the waveguide with respect to the crystallographic axes and transform the components of the relevant vector and tensor quantities to the crystal reference frame. Let $\hat{\mathfrak{e}}_k$, $k=1,2,3$ be the unit vectors defining the crystallographic orientations and let $\left(\hat{\mathfrak{e}}_k\right)_\alpha$, $\alpha=T, L, V$ be their components in the laboratory frame. Then the vector and tensor components transform from laboratory to crystal frame via a transformation matrix $S$ such that $S_{\alpha,k} = \left(\hat{\mathfrak{e}}_k\right)_\alpha$, that is,
\begin{equation*}
\epsilon^{\mathrm{(cryst)}} = S^\dag \epsilon^{\mathrm{(lab)}} S,\quad\quad
\hat{\bm{B}}_{\mathrm{MW}}^{\mathrm{(cryst)}} = S^\dag \hat{\bm{B}}_{\mathrm{MW}}^{\mathrm{(lab)}},
\end{equation*}
where $\epsilon^{\mathrm{(cryst)}}$ and $\epsilon^{\mathrm{(lab)}}$ are the matrices of components of the strain tensor in the two reference frames and $\hat{\bm{B}}_{\mathrm{MW}}^{\mathrm{(cryst)}}$ and $\hat{\bm{B}}_{\mathrm{MW}}^{\mathrm{(lab)}}$ are the vectors of components of the MW field.
In our present study, the structure is oriented along the crystallographic axes, so $S$ is the identity matrix.

The second step is to transform to the NV frame for each orientation of the NV centers, $n=1$, 2, 3, 4. For the first orientation, $n=1$, the transformation matrix is composed of the unit vectors of the NV reference frame, as given in Ref.~\cite{Udvarhelyi2018Spin-strainDiamond}, as its columns,
\begin{equation}\label{eq:Udvarhelyi-reper0}
T^{(1)} = \left(\begin{array}{ccc}
-1/\sqrt{6} & 1/\sqrt{2} & 1/\sqrt{3} \\
-1/\sqrt{6} & -1/\sqrt{2} & 1/\sqrt{3} \\
\sqrt{2/3} & 0 & 1/\sqrt{3} 
\end{array} \right).
\end{equation}
The other three transformation matrices can be constructed by rotating this reference frame to the three remaining NV orientations, $T^{(n)}=R^{(n)}T^{(1)}$, where $R^{(n)}$ are rotation matrices for rotations by $\pi$ with respect to the axes $X,Y,Z$ for $n=2,3,4$, respectively. This results in
\begin{align*}
T^{(2)} &= \left(\begin{array}{ccc}
-1/\sqrt{6} & 1/\sqrt{2} & 1/\sqrt{3} \\
1/\sqrt{6} & 1/\sqrt{2} & -1/\sqrt{3} \\
-\sqrt{2/3} & 0 & -1/\sqrt{3} 
\end{array} \right), \\
T^{(3)} &= \left(\begin{array}{ccc}
1/\sqrt{6} & -1/\sqrt{2} & -1/\sqrt{3} \\
-1/\sqrt{6} & -1/\sqrt{2} & 1/\sqrt{3} \\
-\sqrt{2/3} & 0 & -1/\sqrt{3} 
\end{array} \right), \\
T^{(4)} &= \left(\begin{array}{ccc}
1/\sqrt{6} & -1/\sqrt{2} & -1/\sqrt{3} \\
1/\sqrt{6} & 1/\sqrt{2} & -1/\sqrt{3} \\
\sqrt{2/3} & 0 & 1/\sqrt{3} 
\end{array} \right).
\end{align*}
The components of the strain tensor and MW field are then transformed to the NV reference frames according to
\begin{equation*}
\epsilon^{\mathrm{(n)}} = T^{(n)\dag} \epsilon^{\mathrm{(cryst)}} T^{(n)},\quad\quad
\hat{\bm{B}}_{\mathrm{MW}}^{\mathrm{(n)}} 
= T^{(n)\dag} \hat{\bm{B}}_{\mathrm{MW}}^{\mathrm{(cryst)}}.
\end{equation*}

\section{\MakeUppercase{Stress profile across the waveguide}}
\label{appendix:stress}

Upon fitting the ODMR data, we extracted the strain profile across the waveguide as described in Sec.~\ref{sec:results}. Here, we calculate the corresponding stress profile across the waveguide using Hook's law. Due to the Poisson effect, strain in the transverse direction induces stress in the axial (longitudinal) direction. Thus, for three strain components, there are four non-zero stress components:
\begin{subequations}\label{eqn:stress_components}
\begin{align}
    \sigma_{TT} &= C_{11} \epsilon_{TT} + C_{12}\epsilon_{VV},\\
    \sigma_{LL} &=C_{12}\epsilon_{TT} + C_{12} \epsilon_{VV}, \\
    \sigma_{VV} &= C_{12}\epsilon_{TT} + C_{11}\epsilon_{VV}, \\
    \sigma_{TV} &= C_{44}\epsilon_{TV},
\end{align}
\end{subequations}
where $C_{11}=1076$~GPa, $C_{12}=125$~GPa, and $C_{44}=576$~GPa are the diamond stiffness constants \cite{Kaxiras2003Atomic}.

\begin{figure}[tb]
    \centering
    \includegraphics[width =\linewidth]{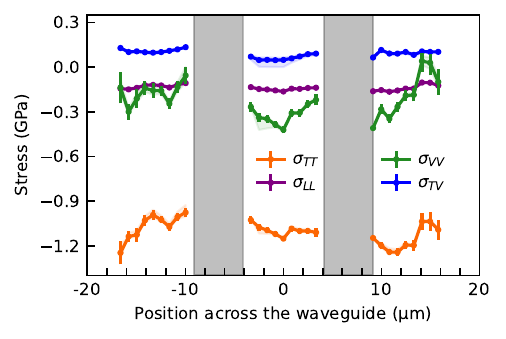}
    \caption{Stress profiles across the waveguide. Individual stress tensor elements plotted versus the position: transverse \( \sigma_{TT} \) (orange), longitudinal \( \sigma_{LL} \) (purple), vertical \( \sigma_{VV} \) (green), and the only nonzero shear component \( \sigma_{TV} \) (blue), which is insensitive to its sign, so we plot its absolute value. Shaded regions represent the uncertainties due to the Lorentzian width ($\gamma$).}
    \label{fig:stress}
\end{figure}

Fig.~\ref{fig:stress} shows the four stress components across the waveguide calculated using Eq.~\eqref{eqn:stress_components}. The stress within the waveguide is dominated by the transverse component (orange curve, \(\sigma_{TT}\)), which is compressive both inside and outside the waveguide. The purple curve shows the longitudinal stress component \( \sigma_{LL} \), which is weak, compressive, and virtually constant across the structures. The green curve represents the vertical stress component (\(\sigma_{VV}\)), which is also compressive and varies regularly within the waveguide, reaching the highest magnitude in its center. The blue curve represents the shear stress component (\(\sigma_{TV}\)), for which we show absolute values, as ODMR spectra are insensitive to its sign.

\end{appendix}

\FloatBarrier

\end{document}